\documentclass[aps,prd,preprint,preprintnumbers]{revtex4}
\usepackage{ae}
\usepackage{bm} 
\usepackage{amssymb}
\usepackage{amsmath}
\usepackage{amsfonts}
\usepackage{graphicx}
\usepackage{slashed}
\usepackage{setspace}
\usepackage{subfigure}
\usepackage{ulem}
\usepackage{bbold}

\newcommand{\Eqref}[1]{Eq.~\eqref{#1}}

\allowdisplaybreaks

\begin{document}

\title{Towards the full Heisenberg-Euler effective action at large $N$}

\author{Felix Karbstein}\email{felix.karbstein@uni-jena.de}
\affiliation{Helmholtz-Institut Jena, Fr\"obelstieg 3, 07743 Jena, Germany}
\affiliation{GSI Helmholtzzentrum f\"ur Schwerionenforschung, Planckstra\ss e 1, 64291 Darmstadt}
\affiliation{Theoretisch-Physikalisches Institut, Abbe Center of Photonics, \\ Friedrich-Schiller-Universit\"at Jena, Max-Wien-Platz 1, 07743 Jena, Germany}

\date{\today}

\begin{abstract}
We study the Heisenberg-Euler effective action in constant electromagnetic fields $\bar{F}$ for QED with $N$ charged particle flavors of the same mass and charge $e$ in the large $N$ limit characterized by sending $N\to\infty$ while keeping $Ne^2\sim e\bar{F}\sim N^0$ fixed. This immediately implies that contributions that scale with inverse powers of $N$ can be neglected and the resulting effective action scales linearly with $N$.
Interestingly, due to the presence of one-particle reducible diagrams, even in this limit the Heisenberg-Euler effective action receives contributions of arbitrary loop order.
In particular for the special cases of electric- and magnetic-like field configurations we construct an explicit expression for the associated effective Lagrangian that, upon extremization for two constant scalar coefficients, allows to evaluate its full, all-order result at arbitrarily large field strengths. We demonstrate that our manifestly nonperturbative expression correctly reproduces the known results for the Heisenberg-Euler effective action at large $N$, namely its all-loop strong field limit and its low-order perturbative expansion in powers of the fine-structure constant. 
\end{abstract}

\maketitle

\section{Introduction}\label{sec:intro}

In a previous article \cite{Karbstein:2021gdi} we advocated the study of external-field quantum electrodynamics (QED) with $N$ charged particle flavors of the same mass $m$ and charge $e$ in the large $N$ limit. 
We emphasized that this deformation of standard QED constitutes a very interesting theoretical laboratory allowing to assess the impact of high-order loop corrections on the physics predictions of a quantum field theory under extreme conditions, and detailed the structure of the associated Heisenberg-Euler effective action $\Gamma_{\rm HE}$ governing the dynamics of prescribed macroscopic electromagnetic fields in the quantum vacuum \cite{Heisenberg:1935qt,Weisskopf:1996bu,Schwinger:1951nm}; see Refs.~\cite{Dunne:2004nc,Fedotov:2022ely} for reviews.
The explicit results of that work mainly assumed the possibility of a perturbative expansion of  $\Gamma_{\rm HE}$ in powers of the fine-structure constant $\alpha=e^2/(4\pi)$, or equivalently, the numbers of loops of its constituting Feynman diagrams.
Note that an $\ell$-loop contribution to $\Gamma_{\rm HE}$  scales as $\alpha^{\ell-1}$.
In our notation, $\alpha\equiv\alpha(\mu^2=m^2)$ and $e\equiv e(\mu^2=m^2)$ are the low-energy values of the fine-structure constant and charge evaluated at the momentum scale $\mu$ set by the mass $m$ of the charged particles, respectively. 

The starting point of our present work is the explicit expression for $\Gamma_{\rm HE}$ derived in Ref.~\cite{Karbstein:2021gdi}, which is a functional of both the gauge potential of the prescribed external electromagnetic field $\bar A^\mu$ and an auxiliary, infrared (IR) divergent vector field $\bar j^\mu$.
By extremization for $\bar j^\mu$ this in principle allows to determine the full result for $\Gamma_{\rm HE}[\bar A]$ for large $N$ QED in the 't Hooft limit \cite{tHooft:1973alw}, characterized by sending $N\to\infty$ while keeping $N\alpha$ and $e\bar A^\mu$ fixed.
We in particular show that the extremization of the associated constant-field Lagrangian ${\cal L}_{\rm HE}\big|_{N\to\infty}$ for $\bar j^\mu$ can be traded for the much simpler extremization for four constant scalar coefficients.
Remarkably, the resulting expression for ${\cal L}_{\rm HE}\big|_{N\to\infty}$ is fully determined by the renowned one-loop Lagrangian ${\cal L}_{\rm HE}^{1\text{-loop}}$ originally derived by Heisenberg and Euler \cite{Heisenberg:1935qt}.
It can be represented entirely in terms of a parameter integral involving ${\cal L}_{\rm HE}^{1\text{-loop}}$ and its two lowest-order derivatives for the field strength tensor $\bar F^{\mu\nu}=\partial^\mu\bar A^\nu-\partial^\nu\bar A^\mu$.
In the special case of electric- and magnetic-like field configurations characterized by the vanishing of one of the secular invariants of the electromagnetic field the situation becomes especially simple and an extremization for just two scalar parameters is required.
Because our approach does not resort to any perturbative expansion, it genuinely grants access to the fully nonperturbative parameter regime.
This is in particular relevant because, though being of genuine theoretical interest, a controlled exploration of this regime is currently out of reach for ordinary one-flavor QED and has not been studied so far.
Its study will hence facilitate unprecedented insights, and -- because of sharing important aspects with standard $N=1$ QED -- will also provide some guidance about the strong-field behavior to be expected for physically realized external-field QED with $N=1$.

Our article is organized as follows: In Sec.~\ref{sec:HE} we construct an explicit expression for the full Heisenberg-Euler effective Lagrangian in the large $N$ limit that, apart from the constant external electromagnetic field $\bar F^{\mu\nu}$, depends on an auxiliary, IR divergent vector field $\bar j^\mu$ which is supposed to extremize this Lagrangian. We show that the problem can be reformulated in a way allowing to trade the rather complicated extremization for $\bar j^\mu$ for an extremization with respect to four constant scalar coefficients. In Sec.~\ref{sec:G=0} we then focus on the special case of electric- and magnetic-like field configurations.
In this limit, additional simplifications are possible and an extremization of the effective Lagrangian for just two scalar constant coefficients is needed. We perform two independent consistency checks of our result for the effective Lagrangian in an electric- or magnetic-like external field, namely reproduce (A) the all-loop expression previously obtained in Ref.~\cite{Karbstein:2019wmj}, and (B) the perturbative result up to ${\cal O}(\alpha^3)$ \cite{Karbstein:2021gdi}. Finally, we end with conclusions and a brief outlook in Sec.~\ref{sec:concls}.

\section{Towards the effective Lagrangian in constant fields}\label{sec:HE}

Our starting point is the formal mean-field expression of $\Gamma_{\rm HE}$ for $N$-flavor QED given in Eq.~(3.6) of Ref.~\cite{Karbstein:2021gdi},
\begin{equation}
 \Gamma_{\rm HE}[\bar A]\big|_{N\to\infty}=-\frac{1}{4}\int\bar F_{\mu\nu}\bar F^{\mu\nu}+S_\psi[\bar A+\bar j]-\frac{1}{2}
\iint \bar{j}_\mu(D^{-1})^{\mu\nu}\bar{j}_\nu\,, 
\label{eq:largeNGamma}
\end{equation}
where $S_\psi[A]=-{\rm i}N\ln\det(-{\rm i}\slashed{D}[A]+m)$, $(D^{-1})^{\mu\nu}$ is the free inverse photon propagator and the auxiliary field $\bar j^\mu$ is implicitly defined in terms of $\bar A^\mu$ by the condition
\begin{equation}
\frac{\delta}{\delta \bar{j}_\mu}	\Gamma_{\rm HE}[\bar A]\big|_{N\to\infty}=0\,.
\label{eq:jintermsofbarA}
\end{equation}
Throughout this work, we use the shorthand notations $\int_x\equiv\int{\rm d}^4x$ and $\int_p\equiv\int{\rm d}^4p/(2\pi)^4$ for integrations over position and momentum space, respectively, and $\int$ for integrations that can be performed either in position or momentum space.

Presuming that both $N\alpha$ and $e\bar A^\mu$ scale as $\sim N^0$, this result neglects corrections of ${\cal O}(1/N)$ and thus becomes exact for $N\to\infty$; see Ref~\cite{Karbstein:2021gdi} for the details.
It should be noted that the combination $e\bar A^\mu$ amounts to a renormalization group (RG)  invariant and thus can be evaluated at any RG scale $\mu$.
We emphasize that Eqs.~\eqref{eq:largeNGamma} and \eqref{eq:jintermsofbarA} hold for completely generic, i.e., also manifestly inhomogeneous, electromagnetic fields.
Also note that they immediately imply that $\bar j^\mu$ can be formally represented as \cite{Karbstein:2021gdi}
\begin{align}
\bar{j}_{\mu} = \int D_{\mu\sigma}\sum_{n=0} ^\infty\frac{1}{n!}\int\ldots\int \bigl(S^{(n+1)}_\psi[\bar A]\bigr)^{\sigma\sigma_1\ldots\sigma_n} \,\bar{j}_{\sigma_1}\ldots \bar{j}_{\sigma_n} \,,\label{eq:currentlargeNexplfinal}
\end{align}
where $D_{\mu\sigma}$ denotes the free photon propagator, which in momentum space scales as $D_{\mu\sigma}(p)\sim1/p^2$, and we made use of the definition
\begin{equation}
 \big(S_\psi^{(n)}[\bar A]\big)^{\sigma_1\ldots\sigma_n}:= \frac{\delta^n S_\psi[\bar A]}{\delta\bar A_{\sigma_1}\ldots\delta\bar A_{\sigma_n}}\,.
  \label{eq:Spsihochn}
\end{equation}
In the remainder of this work we exclusively focus on constant electromagnetic fields, $\bar F^{\mu\nu}={\rm const.}\,\leftrightarrow\,\partial^\alpha\bar F^{\mu\nu}=0$. In this case, we have $S_\psi[\bar A]=\int_x{\cal L}_{\rm HE}^{1\text{-loop}}(\bar F)$ and the associated Lagrangian is given by \cite{Heisenberg:1935qt,Weisskopf:1996bu,Schwinger:1951nm}
\begin{equation}
 {\cal L}_{\rm HE}^{1\text{-loop}}(\bar F)=-\frac{1}{8\pi^2}\int_{0}^{\infty}\frac{{\rm d}T}{T^3}\,{\rm e}^{-m^2T}\,\biggl(\frac{(ec_+T)(ec_-T)}{\tanh(ec_+T)\tan(ec_-T)}-1-\frac{(ec_+T)^2-(ec_-T)^2}{3}\biggr)\,, \label{eq:L_HE_1loop}
\end{equation}
where
\begin{equation}
c_\pm=\bigl(\sqrt{{\cal F}^2+{\cal G}^2}\pm{\cal F}\bigr)^{1/2}\,,\quad\text{with}\quad{\cal F}=\frac{1}{4}\bar F_{\mu\nu}\bar F^{\mu\nu}\quad\text{and\quad}{\cal G}=\frac{1}{4}\bar F_{\mu\nu}{}^\star\!\bar F^{\mu\nu}\,.
\label{eq:cpm}
\end{equation}

In turn, in constant fields \Eqref{eq:Spsihochn} in momentum space can be compactly represented as
\begin{equation}
 \bigl(S^{(n)}_\psi[\bar A]\bigr)^{\sigma_1\ldots\sigma_n}(p_1,\ldots, p_n)  =(2{\rm i})^n (2\pi)^4\,\delta\Bigl(\sum_{j=1}^n p_j\Bigr)\prod_{j=1}^n\biggl(p_j^{\mu_j} 
 \frac{\partial}{\partial\bar F^{\mu_j}_{\ \ \sigma_j}}+{\cal O}\bigl(p_j^3\bigr)\biggr) {\cal L}^{1\text{-loop}}_\text{HE}(\bar F)\,,\label{eq:Sn_constfield}
\end{equation}
with ${\cal O}\bigl(p_j^3\bigr)\sim p_j^\alpha p_j^\beta p_j^\gamma$ denoting a contribution cubic in $p_j^\mu$ \cite{Gies:2016yaa}.
Our conventions for the Fourier transform are $\bar A^\mu(x)=\int_p{\rm e}^{{\rm i}px}\,\bar A^\mu(p)$.
Accounting for the fact that $\int(D^{-1})^{\mu\nu}\bar j_\nu$ has to vanish identically when considered as an individual object because constant fields cannot supply momentum to charged-particle loops and that in momentum space $\int_p p^\alpha\bar j^\mu(p)={\rm finite}$ \cite{Gies:2016yaa}, the above expressions immediately constrain the momentum dependence of $\bar j^\mu$ to be of the form $\bar j^\mu(p)\sim f(|p|)\,\delta(p)\,p^\nu/p^2$.
Here, $f(|p|)$ denotes a function even in $p^\mu$ that fulfills $f(|cp|)=f(|p|)$ for $c\in\mathbb{R}$ and thus remains finite for $p^\mu\to0$.
Hence, in this case we have
\begin{align}
 \int\ldots\int \bigl(S^{(n+1)}_\psi[\bar A]\bigr)^{\sigma\sigma_1\ldots\sigma_n} \,\bar{j}_{\sigma_1}\ldots \bar{j}_{\sigma_n} 
 &\sim p^\nu\int_{p_1}\ldots\int_{p_n}\delta\Bigl(p+\sum_{j=1}^n p_j\Bigr)\prod_{j=1}^n\delta(p_j)f(|p_j|)\,,
 \label{eq:effintjhochn}
\intertext{where we made use of the fact that $f(|p|)\,p^\alpha p^\beta/p^2\to f(|p|)$ is again an even function of $p^\mu$ with the above property.
Subsequently performing $n-1$ out of the $n$ momentum integrations in \Eqref{eq:effintjhochn}, for $n\geq1$ we find}
&\sim p^\nu\,[f(0)]^{n-1}\int_{p'}\delta(p+p')\,\delta(p')f(|p'|) \nonumber\\
&\sim p^\nu\,[f(0)]^{n-1}\,\delta(p)f(|p|)\,,
\label{eq:effintjhochn1}
\end{align}
with $f(0)=\int_p(2\pi)^4\delta(p)\,f(|p|)$.
On the other hand, for $n=0$ we infer $ \bigl(S^{(1)}_\psi[\bar A]\bigr)^\sigma\sim p^\nu\,\delta(p)$.

Equation~\eqref{eq:effintjhochn1} implies that in each summand that couples $n$ factors of $\bar j^\mu$ on the right hand side of \Eqref{eq:currentlargeNexplfinal} only the momentum dependence $f(|p|)$ of one factor is inherited by the $\bar j^\mu$ on its left hand side.
Therefore, with regard to momentum transfer the right hand side of \Eqref{eq:currentlargeNexplfinal} effectively appears as linear in $\bar j^\mu(p)$.
Hence, schematically we have
\begin{equation}
 \bar j_\mu(p)=D_{\mu\sigma}(p)\Bigl(\bar j_0(p)+\Pi(p)\bar j(p)\Bigr)^\sigma\,,
 \label{eq:jimpl}
\end{equation}
where $\bar j_0^\sigma=\bigl(S^{(1)}_\psi[\bar A]\bigr)^\sigma$, and the 'polarization tensor' $\Pi^{\mu\nu}$ implicitly accounts for all the other couplings to $\bar j^\mu$ that do not exchange momentum with the charged particle loop from the outset, i.e., contribute a factor of $f(0)$ in \Eqref{eq:effintjhochn1}.
Equation~\eqref{eq:jimpl} can be readily solved for $\bar j^\mu(p)$, yielding
\begin{equation}
 \bar j_\mu(p)=\Bigl[\bigl(D^{-1}(p)-\Pi(p)\bigr)^{-1}\Bigr]_{\mu\sigma}\,\bar j_0^\sigma(p)\,.
 \label{eq:jexpl}
\end{equation}
Moreover, in constant fields we clearly have ${\cal L}^{1\text{-loop}}_\text{HE}(\bar F)={\cal L}^{1\text{-loop}}_\text{HE}({\cal F},{\cal G})$ [see \Eqref{eq:L_HE_1loop}], which immediately implies that [cf. \Eqref{eq:Sn_constfield}]
\begin{equation}
 \bar j_0^\sigma(p)={\rm i}(2\pi)^4\delta(p)\,p_\nu\Bigl(\bar F^{\nu\sigma}\frac{\partial}{\partial{\cal F}}+{}^\star\!\bar F^{\nu\sigma}\frac{\partial}{\partial{\cal G}}\Bigr){\cal L}^{1\text{-loop}}_\text{HE}({\cal F},{\cal G})\,.
 \label{eq:j0}
\end{equation}
To be precise, ${\cal L}^{1\text{-loop}}_\text{HE}({\cal F},{\cal G})$ is invariant under parity transformations and thus even in $\cal G$, which transforms as a pseudoscalar.
Because of $g^{\mu\nu}=P_T^{\mu\nu}+P_L^{\mu\nu}$ with transverse $P_T^{\mu\nu}$ and longitudinal $P_L^{\mu\nu}$ projectors, and $P_L^{\mu\nu}=p^\mu p^\nu/p^2$ in momentum space, \Eqref{eq:j0} allows us to infer that $(P_L\,\bar j_0)^\mu=0$ and thus $\bar j_0^\mu=(P_T\bar j_0)^\mu$.
In turn, \Eqref{eq:jexpl} can also be expressed as
\begin{equation}
 \bar j_\mu(p)=\Bigl[\bigl(P_T\bigl(D^{-1}(p)-\Pi(p)\bigr)\bigr)^{-1}\Bigr]_{\mu\sigma}\,\bar j_0^\sigma(p)\,.
 \label{eq:jexpl1}
\end{equation}

Note that $\Pi^{\mu\nu}$ has to be quadratic in the momentum transfer $\Pi^{\mu\nu}(p)\sim p^\alpha p^\beta$ in order for \Eqref{eq:jexpl} to be compatible with the momentum structure of $\bar j^\mu$ established above, which is substantially more involved than originally foreseen in Ref.~\cite{Karbstein:2021gdi}; cf. in particular the discussion in context of Eq.~(3.16) therein.

The Ward identity $p_\mu\Pi^{\mu\nu}(p)=\Pi^{\mu\nu}(p)p_\nu=0$ immediately constrains the relevant contribution to $\Pi^{\mu\nu}(p)$ quadratic in $p^\alpha$ to be of the following form \cite{Karbstein:2015cpa},
\begin{align}
\Pi^{\mu\nu}(p)=\pi_T p^2P_T^{\mu\nu}&+\pi_{\bar F\bar F}(p\bar F)^\mu(p\bar F)^\nu+\pi_{{}^\star\!\bar F{}^\star\!\bar F}(p{}^\star\!\bar F)^\mu(p{}^\star\!\bar F)^\nu \nonumber\\
&+\pi_{{}^\star\!\bar F\bar F}\bigl[(p{}^\star\!\bar F)^\mu(p\bar F)^\nu+(p\bar F)^\mu(p{}^\star\!\bar F)^\nu\bigr]\,,
\label{eq:Pi}
\end{align}
where the coefficients $\pi_p$ with $p\in\{T,\bar F\bar F,{}^\star\!\bar F{}^\star\!\bar F,{}^\star\!\bar F\bar F \}$ are functions of both $\cal F$ and $\cal G$.
We emphasize that parity invariance of QED implies the scalars $\pi_T$, $\pi_{\bar F\bar F}$, $\pi_{\bar F{}^\star\!\bar F}$ to be even and $\pi_{{}^\star\!\bar F\bar F}$ to be odd in $\cal G$.
Upon introducing the four-vectors
\begin{equation}
 v_{\parallel/\perp}^\mu=\frac{c_\pm(p{}^\star\!\bar F)^\mu\mp c_\mp(p\bar F)^\mu}{c_+^2+c_-^2}\,,
 \label{eq:vs}
\end{equation}
we define $P_\parallel^{\mu\nu}=v_\parallel^\mu v_\parallel^\nu/v_\parallel^2$, $P_\parallel^{\mu\nu}=v_\perp^\mu v_\perp^\nu/v_\perp^2$, $P_0^{\mu\nu}=(P_T-P_\parallel-P_\perp)^{\mu\nu}$ as well as $Q^{\mu\nu}=v_\parallel^\mu v_\perp^\nu+v_\perp^\mu v_\parallel^\nu$  \cite{Dittrich:2000zu}.
It is easy to verify that for ${\cal G}\geq0$ the tensors $P_0^{\mu\nu}$,  $P_\parallel^{\mu\nu}$ and $P_\perp^{\mu\nu}$ are mutually orthogonal projectors.
With these definitions \Eqref{eq:Pi} can alternatively be represented as
\begin{equation}
 \Pi^{\mu\nu}(p)=\Pi_0P_0^{\mu\nu}+\Pi_\parallel P_\parallel^{\mu\nu}+\Pi_\perp P_\perp^{\mu\nu}+\pi_Q Q^{\mu\nu}\,,
 \label{eq:Pialt}
\end{equation}
where
\begin{align}
\Pi_0&=p^2\pi_T\,, \nonumber\\
\Pi_\parallel&=(c_-^2\pi_{\bar F\bar F}+c_+^2\pi_{{}^\star\!\bar F{}^\star\!\bar F}-2c_+c_-\pi_{{}^\star\!\bar F\bar F}) v_\parallel^2+\pi_T p^2\,, \nonumber\\
\Pi_\perp&=(c_+^2\pi_{\bar F\bar F}+c_-^2\pi_{{}^\star\!\bar F{}^\star\!\bar F}+2c_+c_-\pi_{{}^\star\!\bar F\bar F}) v_\perp^2+\pi_T p^2\,, \nonumber
\end{align}
and $\pi_Q=(c_+^2-c_-^2)\pi_{{}^\star\!\bar F\bar F}-c_+c_-(\pi_{\bar F\bar F}-\pi_{{}^\star\!\bar F{}^\star\!\bar F})$.
Also note that $\bigl(P_T\Pi(p)\bigr)^{\mu\nu}=\Pi^{\mu\nu}(p)$ and $\bigl(P_TD^{-1}(p)\bigr)^{\mu\nu}=p^2P_T^{\mu\nu}$.
In turn, we can write
\begin{align}
\big[P_T\bigl(D^{-1}(p)-\Pi(p)\bigr)\bigr]^{\mu\nu}=(p^2-\Pi_0)P_0^{\mu\nu}&+(p^2-\Pi_\parallel)P_\parallel^{\mu\nu} \nonumber\\
&+(p^2-\Pi_\perp)P_\perp^{\mu\nu}-\pi_QQ^{\mu\nu}\,.
\label{eq:invprop}
\end{align}
The analogous result for ${\cal G}<0$ can be represented in the same form; it actually only requires setting $c_+c_-\to-c_+c_-$ in the above expressions.

In a next step, we aim at the inversion of \Eqref{eq:invprop} which constitutes an important input to \Eqref{eq:jexpl1}.
This expression is clearly of the structure
\begin{equation}
 M^{\mu\nu}=aP_0^{\mu\nu}+bP_\parallel^{\mu\nu}+cP_\perp^{\mu\nu}+dQ^{\mu\nu}\,,
\end{equation}
which can be readily inverted, yielding
\begin{equation}
 (M^{-1})^{\mu\nu}=\frac{1}{a}P_0^{\mu\nu}+\frac{1}{b-v_\parallel^2v_\perp^2d^2/c}\biggl[P_\parallel^{\mu\nu}+\frac{b}{c}P_\perp^{\mu\nu}-\frac{d}{c}Q^{\mu\nu}\biggr]\,.
\end{equation}

Let us now make use of the above insights into the structure of $\bar j^\mu$ in constant fields in the effective action $\Gamma_{\rm HE}$ as given in \Eqref{eq:largeNGamma}.
The basic idea is that upon parametrizing $\bar j^\mu$ in terms of the unknown scalars $\pi_p$ with $p\in\{T,\bar F\bar F,{}^\star\!\bar F{}^\star\!\bar F,{}^\star\!\bar F\bar F\}$ the extremization of  $\Gamma_{\rm HE}$ for $\bar j^\mu$ in \Eqref{eq:jintermsofbarA} can be traded for the much simpler extremization with respect to the four scalar components $\pi_p$.
In contrast to the four-vector field $\bar j^\mu$ that scales as $\bar j^\mu(cp)\sim\delta(cp)/c$ with $c\in\mathbb{R}$ [cf. above \Eqref{eq:effintjhochn}] and thus is infrared divergent, the associated field strength tensor $\bar J^{\mu\nu}:=\partial^\mu\bar j^\nu-\partial^\nu\bar j^\mu$ is manifestly IR finite.
Hence, in general we find it more convenient to work with it.
Resorting to \Eqref{eq:jexpl1} this field strength tensor can be expressed as
\begin{equation}
 \bar J^{\mu\nu}(x)={\rm i}\bigl(g^{\rho\mu}g^{\sigma\nu}-g^{\rho\nu}g^{\sigma\mu}\bigr)
\int_p{\rm e}^{{\rm i}px}\,p_\rho\Bigl[\bigl(P_T\bigl(D^{-1}(p)-\Pi(p)\bigr)\bigr)^{-1}\Bigr]_{\sigma\tau}\bar j_0^\tau(p)\,.
\label{eq:J}
 \end{equation}
Equation~\eqref{eq:J} clearly fulfills $\partial^\alpha\bar J^{\mu\nu}(x)=0$, such that the field strength tensor $(\bar F+\bar J)^{\mu\nu}$ of the combined field $(\bar A+\bar j)^\mu$ amounts to that of a constant field.
In turn, for $\bar F^{\mu\nu}={\rm const}.$ we can write
\begin{align}
 \Gamma_{\rm HE}[\bar A]\big|_{N\to\infty}&=-\frac{1}{4}\int_x\bar F_{\mu\nu}\bar F^{\mu\nu}+\int_x{\cal L}_{\rm HE}^{1\text{-loop}}(\bar F+\bar J)-\frac{1}{4}\int_x\bar J_{\mu\nu}\bar J^{\mu\nu}\,, 
 \label{eq:largeNGamma1-0}
\intertext{which can also be represented as}
&=-\frac{1}{4}\int_x\bar F_{\mu\nu}\bar F^{\mu\nu}+\int_x{\cal L}_{\rm HE}^{1\text{-loop}}(\bar F)+\int_x\frac{\partial{\cal L}_{\rm HE}^{1\text{-loop}}(\bar F)}{\partial\bar F^{\mu\nu}}\bar J^{\mu\nu}-\frac{1}{4}\int_x\bar J_{\mu\nu}\bar J^{\mu\nu} \nonumber\\
&\quad\quad+\int_x\sum_{n=2}^\infty\frac{1}{n!}\Bigl(\bar J^{\rho\sigma}\frac{\partial}{\partial\bar F^{\rho\sigma}}\Bigr)^{n-2}\,\bar J^{\alpha\beta}\,\frac{\partial^2{\cal L}_{\rm HE}^{1\text{-loop}}(\bar F)}{\partial\bar F^{\alpha\beta}\partial\bar F^{\mu\nu}}\bar J^{\mu\nu}\,,
\label{eq:largeNGamma1}
\end{align}
where the derivatives for the field strength tensor $\bar F^{\mu\nu}$ act exclusively on ${\cal L}_{\rm HE}^{1\text{-loop}}(\bar F)$ and not on $\bar J^{\mu\nu}\equiv\bar J^{\mu\nu}(x)$.

Performing the integration over space-time in the last line of \Eqref{eq:largeNGamma1} the exponential factors encoding the space-time dependence of each factor of $\bar J^{\mu\nu}$ combine to a momentum conserving Dirac delta function.
In line with the discussion in the context of Eqs.~\eqref{eq:effintjhochn} and \eqref{eq:effintjhochn1} above, this immediately implies that $n-2$ out of $n$ factors of $\bar J^{\mu\nu}$ that are effectively coupled together do not exchange any momentum with the charged particle loop from the outset.
Correspondingly, the expression in the last line of \Eqref{eq:largeNGamma1} can be cast in the following form
\begin{equation}
\sum_{n=2}^\infty\frac{1}{n!}\Bigl(\bar J^{\rho\sigma}(0)\frac{\partial}{\partial\bar F^{\rho\sigma}}\Bigr)^{n-2}\int_x\bar J^{\alpha\beta}\,\frac{\partial^2{\cal L}_{\rm HE}^{1\text{-loop}}(\bar F)}{\partial\bar F^{\alpha\beta}\partial\bar F^{\mu\nu}}\bar J^{\mu\nu}\,,
\label{eq:lastline}
\end{equation}
where $\bar J^{\mu\nu}(0)$ is obviously given by \Eqref{eq:J} with ${\rm e}^{{\rm i}px}\to1$.
In a next step, we note that the quantity $\partial^2{\cal L}_{\rm HE}^{1\text{-loop}}(\bar F)/\partial\bar F^{\alpha\beta}\partial\bar F^{\mu\nu}$ evaluated in the combined field $\bigl(\bar F+\bar J(0)\bigr)^{\mu\nu}$ can be expressed as 
\begin{equation}
\frac{\partial^2{\cal L}_{\rm HE}^{1\text{-loop}}(\bar F)}{\partial\bar F^{\alpha\beta}\partial\bar F^{\mu\nu}}\bigg|_{\bar F\to\bar F+\bar J(0)}=\sum_{n=0}^\infty\frac{1}{n!}\Bigl(\bar J^{\rho\sigma}(0)\frac{\partial}{\partial\bar F^{\rho\sigma}}\Bigr)^n\,\frac{\partial^2{\cal L}_{\rm HE}^{1\text{-loop}}(\bar F)}{\partial\bar F^{\alpha\beta}\partial\bar F^{\mu\nu}}\,,
\label{eq:tbc}
\end{equation}
the structure of which -- apart from a factor of $(n+1)(n+2)$ in each summand -- closely resembles the infinite sum occurring in \Eqref{eq:lastline}.
Making use of the following identity
\begin{equation}
\int_0^1{\rm d}K\int_0^K{\rm d}\kappa\,\kappa^n=\frac{1}{n+1}\frac{1}{n+2}\,,
\end{equation}
this mismatch can, however, be eliminated and allows us to establish that
\begin{align}
&\int_0^1{\rm d}K\int_0^K{\rm d}\kappa\,\frac{\partial^2{\cal L}_{\rm HE}^{1\text{-loop}}(\bar F)}{\partial\bar F^{\alpha\beta}\partial\bar F^{\mu\nu}}\bigg|_{\bar F\to\bar F+\kappa\bar J(0)}\nonumber\\
&\hspace*{2cm}=\int_0^1{\rm d}K\int_0^K{\rm d}\kappa\,\sum_{n=0}^\infty\frac{1}{n!}\Bigl(\kappa\bar J^{\rho\sigma}(0)\frac{\partial}{\partial\bar F^{\rho\sigma}}\Bigr)^n\,\frac{\partial^2{\cal L}_{\rm HE}^{1\text{-loop}}(\bar F)}{\partial\bar F^{\alpha\beta}\partial\bar F^{\mu\nu}} \nonumber\\
&\hspace*{2cm}=\sum_{n=2}^\infty\frac{1}{n!}\Bigl(\kappa\bar J^{\rho\sigma}(0)\frac{\partial}{\partial\bar F^{\rho\sigma}}\Bigr)^{n-2}\,\frac{\partial^2{\cal L}_{\rm HE}^{1\text{-loop}}(\bar F)}{\partial\bar F^{\alpha\beta}\partial\bar F^{\mu\nu}}\,,
\label{eq:tbc1}
\end{align}
which upon contraction with $\bar J^{\alpha\beta}\bar J^{\mu\nu}$ and integration over space-time agrees with \Eqref{eq:lastline}.
In turn, the effective action~\eqref{eq:largeNGamma1-0} can be alternatively represented as

\begin{align}
 \Gamma_{\rm HE}[\bar A]\big|_{N\to\infty}&=-\frac{1}{4}\int_x\bar F_{\mu\nu}\bar F^{\mu\nu}+\int_x{\cal L}_{\rm HE}^{1\text{-loop}}(\bar F)+\int_x\frac{\partial{\cal L}_{\rm HE}^{1\text{-loop}}(\bar F)}{\partial\bar F^{\mu\nu}}\bar J^{\mu\nu}\nonumber\\
&\quad\quad-\int_0^1{\rm d}K\int_0^K{\rm d}\kappa\int_x\bar J^{\alpha\beta}\Biggl[\frac{1}{2}g_{\alpha\mu}g_{\beta\nu}-\frac{\partial^2{\cal L}_{\rm HE}^{1\text{-loop}}(\bar F)}{\partial\bar F^{\alpha\beta}\partial\bar F^{\mu\nu}}\bigg|_{\bar F\to\bar F+\kappa\bar J(0)}\Biggr]\bar J^{\mu\nu}\,,
\label{eq:largeNGamma2}
\end{align}
which now is to be extremized for $\pi_p$ with  $p\in\{T,\bar F\bar F,{}^\star\!\bar F{}^\star\!\bar F,{}^\star\!\bar F\bar F \}$.
Equation~\eqref{eq:largeNGamma2} depends on the latter via $\bar J^{\mu\nu}(0)$ and $\bar J^{\mu\nu}(x)$; cf. \Eqref{eq:J}.
Also note that $\int_x\bar J^{\mu\nu}=V^{(4)}\bar J^{\mu\nu}(0)$ with four dimensional space-time volume $V^{(4)}$ due to $\int_x\int_p{\rm e}^{{\rm i}px}f(p)\delta(p)=V^{(4)}(2\pi)^4\int_p f(p)\delta(p)$.
Because of ${\cal L}^{1\text{-loop}}_\text{HE}(\bar F)={\cal L}^{1\text{-loop}}_\text{HE}({\cal F},{\cal G})$ in constant fields, cf. above \Eqref{eq:j0}, we moreover have
\begin{equation}
 \frac{\partial{\cal L}_{\rm HE}^{1\text{-loop}}(\bar F)}{\partial\bar F^{\mu\nu}}=\frac{1}{2}\Bigl(\bar F_{\mu\nu}\frac{\partial}{\partial{\cal F}}+{}^\star\!\bar F_{\mu\nu}\frac{\partial}{\partial{\cal G}}\Bigr){\cal L}^{1\text{-loop}}_\text{HE}({\cal F},{\cal G}) \nonumber
\end{equation}
and
\begin{align}
 \frac{\partial^2{\cal L}_{\rm HE}^{1\text{-loop}}(\bar F)}{\partial \bar F^{\alpha\beta}\partial \bar F^{\mu\nu}}
 =\frac{1}{4}\Biggl[\bigl(g_{\alpha\mu}g_{\beta\nu}-g_{\alpha\nu}g_{\beta\mu}\bigr)\frac{\partial}{\partial{\cal F}}  +  \epsilon_{\mu\nu\alpha\beta}\frac{\partial}{\partial{\cal G}} 
 + \bar F_{\alpha\beta} \bar F_{\mu\nu}\frac{\partial^2}{\partial{\cal F}^2} + {}^\star\!\bar F_{\alpha\beta}{}^\star\!\bar F_{\mu\nu}\frac{\partial^2}{\partial{\cal G}^2} \nonumber\\
 + \bigl({}^\star\!\bar F_{\alpha\beta}\bar F_{\mu\nu} + \bar F_{\alpha\beta}{}^\star\!\bar F_{\mu\nu}\bigr)\,\frac{\partial^2}{\partial{\cal F}\partial{\cal G}}\Biggr]{\cal L}_{\rm HE}^{1\text{-loop}}({\cal F},{\cal G}) \,,
 \label{eq:ddL}
\end{align}
where we explicitly accounted for the (anti)symmetry of the latter expression with regard to its indices \cite{Karbstein:2015cpa}.

With the definitions
\begin{align}
{\cal F}_\kappa&=\frac{1}{4}\bigl(\bar F+\kappa\bar J(0)\bigr)_{\mu\nu}\bigl(\bar F+\kappa\bar J(0)\bigr)^{\mu\nu}\,, \nonumber\\
{\cal G}_\kappa&=\frac{1}{4}\bigl(\bar F+\kappa\bar J(0)\bigr)_{\mu\nu}{}^\star\!\bigl(\bar F+\kappa\bar J(0)\bigr)^{\mu\nu}\,,
\end{align}
fulfilling ${\cal F}_0={\cal F}$ and ${\cal G}_0={\cal G}$, the Lagrangian associated with \Eqref{eq:largeNGamma2} is thus given by
\begin{align}
&{\cal L}_{\rm HE}({\cal F},{\cal G})\big|_{N\to\infty}=-{\cal F}+{\cal L}_{\rm HE}^{1\text{-loop}}({\cal F},{\cal G}) +\frac{1}{2}\Bigl(\bar F_{\mu\nu}\frac{\partial{\cal L}^{1\text{-loop}}_\text{HE}({\cal F},{\cal G}) }{\partial{\cal F}}+{}^\star\!\bar F_{\mu\nu}\frac{\partial{\cal L}^{1\text{-loop}}_\text{HE}({\cal F},{\cal G}) }{\partial{\cal G}}\Bigr)\bar J^{\mu\nu}(0) \nonumber\\
&\hspace*{1cm}-\frac{1}{4}\int_0^1{\rm d}K\int_0^K{\rm d}\kappa\,\biggl\{2\,g_{\alpha\mu}g_{\beta\nu}\Bigl(1-\frac{\partial{\cal L}^{1\text{-loop}}_\text{HE}({\cal F}_\kappa,{\cal G}_\kappa) }{\partial{\cal F}_\kappa}\Bigr) -\epsilon_{\mu\nu\alpha\beta}\frac{\partial{\cal L}^{1\text{-loop}}_\text{HE}({\cal F}_\kappa,{\cal G}_\kappa) }{\partial{\cal G}_\kappa}\nonumber\\
&\hspace*{4.8cm}-\bigl(\bar F+\kappa\bar J(0)\bigr)_{\alpha\beta}\bigl(\bar F+\kappa\bar J(0)\bigr)_{\mu\nu}\frac{\partial^2{\cal L}^{1\text{-loop}}_\text{HE}({\cal F}_\kappa,{\cal G}_\kappa) }{\partial{\cal F}_\kappa^2} \nonumber\\
&\hspace*{4.8cm}-{}^\star\!\bigl(\bar F+\kappa\bar J(0)\bigr)_{\alpha\beta}{}^\star\!\bigl(\bar F+\kappa\bar J(0)\bigr)_{\mu\nu}\frac{\partial^2{\cal L}^{1\text{-loop}}_\text{HE}({\cal F}_\kappa,{\cal G}_\kappa) }{\partial{\cal G}_\kappa^2} \nonumber\\
&\hspace*{4.8cm}-2\,{}^\star\!\bigl(\bar F+\kappa\bar J(0)\bigr)_{\alpha\beta}\bigl(\bar F+\kappa\bar J(0)\bigr)_{\mu\nu}\frac{\partial^2{\cal L}^{1\text{-loop}}_\text{HE}({\cal F}_\kappa,{\cal G}_\kappa) }{\partial{\cal F}_\kappa\partial{\cal G}_\kappa}\biggr\} \nonumber\\
&\hspace*{8cm}\times\frac{1}{V^{(4)}}\int_x \bar J^{\alpha\beta}\bar J^{\mu\nu}\,.
\label{eq:largeNcalL}
\end{align}
The relevant expressions for $\bar J^{\mu\nu}(0)$ and $\int_x \bar J^{\alpha\beta}\bar J^{\mu\nu}$ entering \Eqref{eq:largeNcalL} can be obtained from Eqs.~\eqref{eq:j0} and \eqref{eq:J}. They read
\begin{align}
 \bar J^{\mu\nu}(0)&=\bigl(g^{\tau\mu}g^{\varphi\nu}-g^{\tau\nu}g^{\varphi\mu}\bigr)\Bigl(\bar F^{\rho\sigma}\frac{\partial{\cal L}^{1\text{-loop}}_\text{HE}({\cal F},{\cal G}) }{\partial{\cal F}}+{}^\star\!\bar F^{\rho\sigma}\frac{\partial{\cal L}^{1\text{-loop}}_\text{HE}({\cal F},{\cal G}) }{\partial{\cal G}}\Bigr) \nonumber\\
&\hspace*{1cm}\times\int_p\,p_\tau p_\sigma\Bigl[\bigl(P_T\bigl(D^{-1}(p)-\Pi(p)\bigr)\bigr)^{-1}\Bigr]_{\varphi\rho} (2\pi)^4\delta(p)
\label{eq:J0}
\end{align}
and
\begin{align}
 \frac{1}{V^{(4)}}\int_x \bar J^{\alpha\beta}\bar J^{\mu\nu}&=\bigl(g^{\eta\alpha}g^{\epsilon\beta}-g^{\eta\beta}g^{\epsilon\alpha}\bigr)\bigl(g^{\tau\mu}g^{\varphi\nu}-g^{\tau\nu}g^{\varphi\mu}\bigr) \nonumber\\
&\hspace*{1cm}\times\Bigl(\bar F^{\gamma\delta}\frac{\partial{\cal L}^{1\text{-loop}}_\text{HE}({\cal F},{\cal G}) }{\partial{\cal F}}+{}^\star\!\bar F^{\gamma\delta}\frac{\partial{\cal L}^{1\text{-loop}}_\text{HE}({\cal F},{\cal G}) }{\partial{\cal G}}\Bigr) \nonumber\\
&\hspace*{1cm}\times\Bigl(\bar F^{\rho\sigma}\frac{\partial{\cal L}^{1\text{-loop}}_\text{HE}({\cal F},{\cal G}) }{\partial{\cal F}}+{}^\star\!\bar F^{\rho\sigma}\frac{\partial{\cal L}^{1\text{-loop}}_\text{HE}({\cal F},{\cal G}) }{\partial{\cal G}}\Bigr) \nonumber\\
&\hspace*{1cm}\times\int_p\,p_\eta p_\delta\Bigl[\bigl(P_T\bigl(D^{-1}(p)-\Pi(p)\bigr)\bigr)^{-1}\Bigr]_{\epsilon\gamma} \nonumber\\
&\hspace*{3cm}\times p_\tau p_\sigma\Bigl[\bigl(P_T\bigl(D^{-1}(p)-\Pi(p)\bigr)\bigr)^{-1}\Bigr]_{\varphi\rho}(2\pi)^4\delta(p)\,.
\label{eq:intJJ}
\end{align}
Clearly, the only nontrivial task  in evaluating Eqs.~\eqref{eq:J0} and \eqref{eq:intJJ} is to perform the manifestly finite momentum integrals.

\section{Magnetic- and electric-like fields}\label{sec:G=0}

In the remainder of this work we focus exclusively on the specific situation where ${\cal G}=0$ and only ${\cal F}$ is nonvanishing; ${\cal L}_\text{HE}({\cal F})={\cal L}_\text{HE}({\cal F},{\cal G}=0)$.
This provides access to magnetic-like (${\cal F}>0$) and electric-like (${\cal F}<0$) field configurations, that can be mapped on purely magnetic and electric fields, respectively, by Lorentz transformations.
Without loss of generality, here we focus on the case where ${\cal F}\geq0$.
The complementary regime where ${\cal F}<0$ is accessible therefrom by an analytic continuation $\sqrt{2{\cal F}}\to-{\rm i}\sqrt{-2{\cal F}}$; cf., e.g., Ref.~\cite{Karbstein:2021gdi}.
Correspondingly, Eqs.~\eqref{eq:cpm} and \eqref{eq:vs} imply that for ${\cal F}\geq0$ as considered here we have $c_+=\sqrt{2{\cal F}}$, $c_-=0$ and $v_\parallel^\mu=(p{}^\star\!\bar F)^\mu/\sqrt{2{\cal F}}$ as well as $v_\perp^\mu=(p\bar F)^\mu/\sqrt{2{\cal F}}$.
Especially because the dependence of ${\cal L}^{1\text{-loop}}_\text{HE}({\cal F},{\cal G})$ on $\cal G$ is even, i.e., in terms of ${\cal G}^2$, in this case the expression for ${\cal L}_\text{HE}$ in \Eqref{eq:largeNcalL} simplifies significantly.
Besides, many terms vanish due to the identity $\bar F^{\mu\alpha}{}^\star\!\bar F^\nu_{\,\ \alpha}={}^\star\!\bar F^{\mu\alpha}\bar F^\nu_{\,\ \alpha}={\cal G}g^{\mu\nu}\to0$.
Based on these insights it is straightforward to see that for ${\cal G}=0$ \Eqref{eq:J0} reduces to the much more compact expression
\begin{align}
 \bar J^{\mu\nu}(0)&=\frac{\partial{\cal L}^{1\text{-loop}}_\text{HE}({\cal F}) }{\partial{\cal F}}\int_p\frac{(p\bar F)^\mu p^\nu-(p\bar F)^\nu p^\mu}{(1-\pi_T)p^2-\pi_{\bar F\bar F}(p\bar F)^2}\, (2\pi)^4\delta(p)\,,
\label{eq:J0F}
\end{align}
where we already inserted the explicit result for the inverse of \Eqref{eq:invprop} specialized to ${\cal G}=0$.
Finally, we aim at carrying out the momentum integration in \Eqref{eq:J0F}.
Achieving this requires several steps (i)-(iii): (i) We note the following identity \cite{Karbstein:2023}
\begin{equation}
 \int_p\frac{p^{\sigma_1}\cdots p^{\sigma_{2n}}}{(p^2)^{n}}(2\pi)^d\delta(p)=\frac{g^{(\sigma_1\sigma_2}g^{\sigma_3\sigma_4}\cdots g^{\sigma_{2n-1}\sigma_{2n})}}{d(d+2)(d+4)\cdots(d+2n-2)}\,, \label{eq:momintdelta3}
\end{equation}
in $d$ space-time dimensions, where $g_{\perp,\,\sigma_1\sigma_2}\cdots g_{\perp,\,\sigma_{2n-1}\sigma_{2n}}$ with $g^{(\mu\nu)}=g^{\mu\nu}$ denotes a normalized symmetrization.
(ii) We account for the fact that in a reference system where $2{\cal F}=\vec{B}^2$, i.e., is purely magnetic, we have $(p\bar F)^2|_{{\cal G}=0}=2{\cal F}p_\mu p_\nu g_\perp^{\mu\nu}$, with the metric $g_\perp^{\mu\nu}$ singling out the $d-2$ spatial components perpendicular to the magnetic field $\vec{B}$.
Together (i) and (ii) imply that
\begin{equation}
 \int_p\biggl(\frac{(p\bar F)^2}{p^2}\biggr)^n\bigg|_{{\cal G}=0}(2\pi)^4\delta(p)=\frac{(2{\cal F})^n}{n+1}\,.
\end{equation}
(iii) We rewrite \Eqref{eq:J0F} as
\begin{align}
 \bar J^{\mu\nu}(0)&=\frac{1}{2}\frac{\partial{\cal L}^{1\text{-loop}}_\text{HE}({\cal F})}{\partial{\cal F}}\frac{1}{1-\pi_T}\sum_{n=0}^\infty\frac{1}{n+1}\biggl(\frac{\pi_{\bar F\bar F}}{1-\pi_T}\biggr)^n \nonumber\\
 &\hspace*{2cm}\times\biggl(\frac{\partial}{\partial \bar F_{\nu\mu}}-\frac{\partial}{\partial \bar F_{\mu\nu}}\biggr)\int_p\biggl(\frac{(p\bar F)^2}{p^2}\biggr)^{n+1}\bigg|_{{\cal G}=0}(2\pi)^4\delta(p) \nonumber\\
 &=-\frac{\partial{\cal L}^{1\text{-loop}}_\text{HE}({\cal F})}{\partial{\cal F}}\frac{1}{1-\pi_T}\sum_{n=0}^\infty\frac{1}{n+2}\biggl(\frac{2{\cal F}\pi_{\bar F\bar F}}{1-\pi_T}\biggr)^n\bar F^{\mu\nu} 
 =\xi\,\frac{\chi+\ln(1-\chi)}{\chi^2}\,\bar F^{\mu\nu} \,,
\label{eq:J0F1}
\end{align}
with
\begin{equation}
 \xi:=\frac{\partial{\cal L}^{1\text{-loop}}_\text{HE}({\cal F})}{\partial{\cal F}}\frac{1}{1-\pi_T}\quad\text{and}\quad\chi:=\frac{2{\cal F}\pi_{\bar F\bar F}}{1-\pi_T}\,.
 \label{eq:xi+chi}
\end{equation}
Equation~\eqref{eq:J0F1} immediately implies that for ${\cal G}=0$ also ${\cal G}_\kappa=0$.
On the other hand, in this limit the explicit expression for ${\cal F}_\kappa$ is given by
\begin{equation}
 {\cal F}_\kappa=\biggl[1+\kappa\,\xi\,\frac{\chi+\ln(1-\chi)}{\chi^2}\biggr]^2{\cal F}\,.
 \label{eq:Fkappa}
\end{equation}
Besides, along the same lines as in \Eqref{eq:J0F1} one can show that
\begin{equation}
  \frac{1}{V^{(4)}}\int_x \bar J_{\mu\nu}\bar J^{\mu\nu}=2\xi^2\,\frac{\chi+(1-\chi)\ln(1-\chi)}{(1-\chi)\chi^2}\,2{\cal F}
\end{equation}
and
\begin{equation}
  \frac{1}{V^{(4)}}\int_x \bar J^{\alpha\beta}\bar J^{\mu\nu}\bar F_{\alpha\beta}\bar F_{\mu\nu}=4\xi^2\,\frac{-\chi^2+2\chi+2(1-\chi)\ln(1-\chi)}{(1-\chi)\chi^3}\,(2{\cal F})^2\,.
\end{equation}
Using these results in \Eqref{eq:largeNcalL} we finally arrive at a rather compact expression for the effective Lagrangian in field configurations where ${\cal G}=0$, namely
\begin{align}
{\cal L}_{\rm HE}({\cal F})\big|_{N\to\infty}\,=&-{\cal F}+{\cal L}_{\rm HE}^{1\text{-loop}}({\cal F})+2{\cal F}\xi\frac{\partial{\cal L}^{1\text{-loop}}_\text{HE}({\cal F})}{\partial{\cal F}}f_1(\chi)\nonumber\\
 &\hspace*{-2cm}-2{\cal F}\xi^2\int_0^1{\rm d}K\int_0^K{\rm d}\kappa\,\biggl[\Bigl(1-\frac{\partial{\cal L}^{1\text{-loop}}_\text{HE}({\cal F}_\kappa)}{\partial{\cal F}_\kappa}\Bigr)f_2(\chi) -2{\cal F}_\kappa\,\frac{\partial^2{\cal L}^{1\text{-loop}}_\text{HE}({\cal F}_\kappa) }{\partial{\cal F}_\kappa^2}f_3(\chi)\biggr]\,.
\label{eq:largeNcalLB}
\end{align}
Here, we introduced the shorthand notations
\begin{align}
 f_1(\chi)&=\frac{\chi+\ln(1-\chi)}{\chi^2}\,, \nonumber\\
 f_2(\chi)&=\frac{\chi+(1-\chi)\ln(1-\chi)}{(1-\chi)\chi^2}\,, \nonumber\\
 f_3(\chi)&=\frac{-\chi^2+2\chi+2(1-\chi)\ln(1-\chi)}{(1-\chi)\chi^3}\,,
\end{align}
and the relevant Heisenberg-Euler effective Lagrangian at one loop~\eqref{eq:L_HE_1loop} is now given by
\begin{equation}
{\cal L}^{1\text{-loop}}_\text{HE}({\cal F})=-\frac{N}{8\pi^2}\int_0^\infty\frac{{\rm d}T}{T^3}\,{\rm e}^{-m^2T}\biggl(\frac{e\sqrt{2{\cal F}}\,T}{\tanh\bigl(e\sqrt{2{\cal F}}\,T\bigr)}-1-\frac{(e\sqrt{2{\cal F}}\,T)^2}{3}\biggr)\,.
\label{eq:LHE1loopB}
\end{equation}
Equation~\eqref{eq:Fkappa}, which can also be recast as ${\cal F}_\kappa=\bigl[1+\kappa\xi f_1(\chi)\bigr]^2{\cal F}$, implies the identity ${\cal F}_\kappa\,\partial_{{\cal F}_\kappa}={\cal F}\,\partial_{\cal F}$ to hold, where we introduced the shorthand notations $\partial_{\cal F}:=\partial/\partial{\cal F}$ and $\partial_{{\cal F}_\kappa}:=\partial/\partial{\cal F}_\kappa$.
Extremizing \Eqref{eq:largeNcalLB} for $\pi_T$ and $\pi_{\bar F\bar F}$ is equivalent to an extremization for $\xi$ and $\chi$.
Recall that ${\cal F}_\kappa={\cal F}_\kappa(\xi,\chi)$ is a function of these parameters.
We emphasize that the only step  still required to be implemented in order to arrive at the full Heisenberg-Euler effective Lagrangian at large $N$ is to extremize  \Eqref{eq:largeNcalLB} for the two scalar parameters $\xi$ and $\chi$. The structure of  \Eqref{eq:largeNcalLB} suggests that as the one-loop Lagrangian~\eqref{eq:LHE1loopB} can be evaluated numerically for arbitrary values of $\cal F$ \cite{Dittrich:1978fc,Valluri:1982ip}, the same should hold true for the full Heisenberg-Euler effective Lagrangian at large $N$.
The parameter integral over $T$ in \Eqref{eq:LHE1loopB} can even be performed explicitly resulting in a closed-form representation of ${\cal L}^{1\text{-loop}}_\text{HE}({\cal F})$ in terms of the Hurwitz zeta function \cite{Dittrich:1975au}. See also Refs.~\cite{Heyl:1996dt,Karbstein:2015cpa} for analogous closed-form representations of the first and second derivatives of ${\cal L}^{1\text{-loop}}_\text{HE}({\cal F})$ for $\cal F$ entering  \Eqref{eq:largeNcalLB} .

In particular note that for $\kappa=0$ we have ${\cal F}_\kappa\to{\cal F}$, such that when setting $\kappa=0$ in its integrand \Eqref{eq:largeNcalLB} becomes just quadratic in $\xi$.
This immediately implies that the condition $\partial/\partial\xi\,{\cal L}_{\rm HE}({\cal F})\big|_{N\to\infty}=0$ can be readily solved for $\xi$ in this specific case.
Also note that as is obvious from Eqs.~\eqref{eq:J} and \eqref{eq:largeNGamma2}, a contribution to ${\cal L}_{\rm HE}(\bar F)\big|_{N\to\infty}$ that couples $n\geq3$ factors of $\bar{j}_0^\mu$ scales at least as $\alpha^n$. 
Hence, aiming at the extraction of the perturbative result for ${\cal L}_{\rm HE}(\bar F)\big|_{N\to\infty}$ up to order $\alpha^n$ it is sufficient to account only for contributions up to $\kappa^{n-2}$ with $n\geq2$ in the integrand of the  $\kappa$ integral.

\begin{figure}[h]
\includegraphics[width=0.8\textwidth]{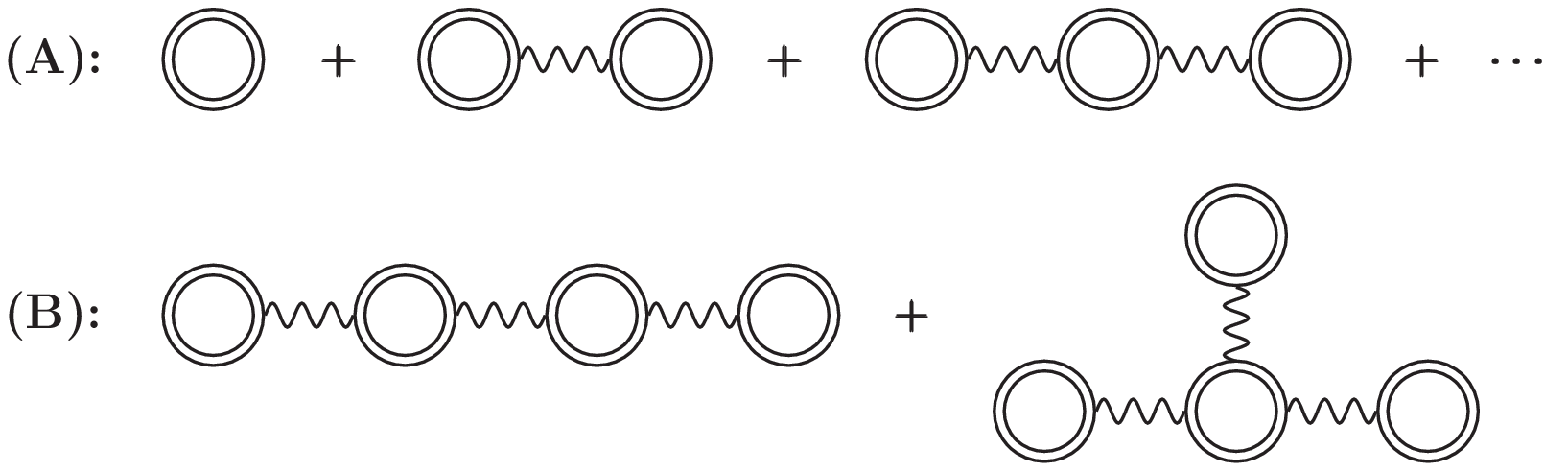}
\caption{{\bf (A):} Infinite set of bubble-chain diagrams constituting  ${\cal L}_{\rm HE}(\bar F)\big|_{N\to\infty}+{\cal F}$ up to quadratic order in $\bar{j}_0^\mu$. The diagrams explicitly depicted here represent the perturbative result up to ${\cal O}(\alpha^2)$. {\bf(B):} Perturbative result at ${\cal O}(\alpha^3)$. The second diagram is cubic in $\bar{j}_0^\mu$.}
\label{fig:diags}
\end{figure}

As nontrivial consistency checks of \Eqref{eq:largeNcalLB} we aim at reproducing (A) the all-order expression obtained when accounting for all possible contributions to ${\cal L}_{\rm HE}({\cal F})\big|_{N\to\infty}$ up to quadratic order in $\bar{j}_0^\mu$ in a direct calculation \cite{Karbstein:2021gdi,Karbstein:2023}, and (B) the perturbative result up to ${\cal O}(\alpha^3)$.
The latter also comprises a contribution cubic in $\bar{j}_0^\mu$. See Fig.~\ref{fig:diags} for an illustration of the relevant Feynman diagrams.

(A): Truncating the Heisenberg-Euler effective Lagrangian at quadratic order in $\bar{j}_0^\mu$ amounts to setting $\kappa=0$ in the integrand of \Eqref{eq:largeNcalLB}.
Correspondingly, we have ${\cal F}_\kappa\to{\cal F}$ and $\int_0^1{\rm d}K\int_0^K{\rm d}\kappa\to1/2$, such that the Lagrangian up to quadratic order in $\bar j_0^\mu$ can be expressed as
\begin{align}
{\cal L}_{\rm HE}({\cal F})\big|_{N\to\infty}^{{\cal O}(\bar j_0^2)}\,=&-{\cal F}+{\cal L}_{\rm HE}^{1\text{-loop}}({\cal F})+2{\cal F}\xi\,\frac{\partial{\cal L}^{1\text{-loop}}_\text{HE}({\cal F})}{\partial{\cal F}}f_1(\chi)\nonumber\\
 &-{\cal F}\xi^2\biggl[\Bigl(1-\frac{\partial{\cal L}^{1\text{-loop}}_\text{HE}({\cal F}) }{\partial{\cal F}}\Bigr)f_2(\chi) -2{\cal F}\,\frac{\partial^2{\cal L}^{1\text{-loop}}_\text{HE}({\cal F}) }{\partial{\cal F}^2}f_3(\chi)\biggr]\,.
\label{eq:largeNcalLBquad}
\end{align}
Solving $\partial/\partial\xi\,{\cal L}_{\rm HE}({\cal F})\big|_{N\to\infty}^{{\cal O}(\bar j_0^2)}=0$ for $\xi$ and plugging this result into  $\partial/\partial\chi\,{\cal L}_{\rm HE}({\cal F})\big|_{N\to\infty}^{{\cal O}(\bar j_0^2)}=0$, the latter equation can be solved for $\chi$.
An explicit solution is found to be given by
\begin{equation}
\chi=\frac{2{\cal F}\,\partial_{\cal F}^2{\cal L}^{1\text{-loop}}_\text{HE}({\cal F})}{1-\partial_{\cal F}{\cal L}^{1\text{-loop}}_\text{HE}({\cal F})}=:\chi_0\,.
\label{eq:chi0}
\end{equation}
Because ${\cal L}^{1\text{-loop}}_\text{HE}({\cal F})$ is a function of $e^2{\cal F}\sim N^0$ only [cf. \Eqref{eq:LHE1loopB}], we clearly have $\partial_{\cal F}{\cal L}^{1\text{-loop}}_\text{HE}({\cal F})\sim{\cal F}\,\partial_{\cal F}^2{\cal L}^{1\text{-loop}}_\text{HE}({\cal F})\sim\alpha$ and $\partial_{\cal F}^2{\cal L}^{1\text{-loop}}_\text{HE}({\cal F})\sim\alpha^2$.
Upon inserting \Eqref{eq:chi0} into \Eqref{eq:largeNcalLBquad}, we obtain
\begin{equation}
{\cal L}_{\rm HE}({\cal F})\big|_{N\to\infty}^{{\cal O}(\bar j_0^2)}\,=-{\cal F}+{\cal L}_{\rm HE}^{1\text{-loop}}({\cal F})-\frac{1}{2}\frac{\bigl(\partial_{\cal F}{\cal L}^{1\text{-loop}}_\text{HE}({\cal F})\bigr)^2}{\partial_{\cal F}^2{\cal L}^{1\text{-loop}}_\text{HE}({\cal F})}\biggl(1+\frac{\ln(1-\chi_0)}{\chi_0}\biggr)\,,
\label{eq:largeNcalLBquad2}
\end{equation}
which matches the corresponding expression determined by a direct calculation in Ref.~\cite{Karbstein:2023}. Making use of the series expansion of  $\ln(1-\chi_0)$ about $\chi_0=0$, yielding $1+\ln(1-\chi_0)/\chi_0=-\sum_{n=1}^\infty\chi_0^n/(n+1)$, it is easy to see that the last term in \Eqref{eq:largeNcalLBquad2} contributes from order $\alpha$, i.e., two loops, onwards.
In particular, with
\begin{equation}
{\cal L}_{\rm HE}^{1\text{-loop}}({\cal F})=e^2{\cal F}\,\frac{\beta_1}{4\pi} \ln\biggl(\frac{e\sqrt{2{\cal F}}}{m^2}\biggr)\,\biggl[1+{\cal O}\biggl(\ln^{-1}\Bigl(\tfrac{e\sqrt{2{\cal F}}}{m^2}\Bigr)\biggr)\biggr]\,,
\end{equation}
where $\beta_1=N/(3\pi)$, and accounting only for the leading strong-field $e\sqrt{2{\cal F}}\gg m^2$ behavior at any given order in a perturbative expansion in $\alpha$, \Eqref{eq:largeNcalLBquad2} reduces to \cite{Karbstein:2019wmj}
\begin{align}
{\cal L}_{\rm HE}({\cal F})\big|_{N\to\infty}^{{\cal O}(\bar j_0^2)}\,=-{\cal F}+e^2{\cal F}\,\frac{\beta_1}{4\pi} \ln\biggl(\frac{e\sqrt{2{\cal F}}}{m^2}\biggr)&\biggl[1+\frac{1}{2} \alpha^{1\text{-loop}}\bigl(e\sqrt{2{\cal F}}\bigr)\beta_1
\ln\biggl(\frac{e\sqrt{2{\cal F}}}{m^2}\biggr)\biggr] \nonumber\\
&\times\biggl[1+{\cal O}\biggl(\ln^{-1}\Bigl(\tfrac{e\sqrt{2{\cal F}}}{m^2}\Bigr)\biggr)\biggr]\,,
\label{eq:largeNcalLBquad3}
\end{align}
with
\begin{equation}
 \alpha^{1\text{-loop}}(\mu^2)=\frac{\alpha}{1-\alpha\beta_1\ln\bigl(\frac{\mu^2}{m^2}\bigr)}\,.
\label{eq:alpha1loop}
\end{equation}

(B): Aiming at extracting the perturbative result for ${\cal L}_{\rm HE}({\cal F})\big|_{N\to\infty}$ up to ${\cal O}(\alpha^3)$ we first recall that  because this receives contributions from diagrams that are at most cubic in $\bar{j}_0^\mu$ for its determination it is sufficient to truncate the integrand of the $\kappa$ integral in \Eqref{eq:largeNcalLB} at linear order in $\kappa$.
Noting that 
\begin{equation}
\frac{\partial{\cal L}^{1\text{-loop}}_\text{HE}({\cal F}_\kappa)}{\partial{\cal F}_\kappa}=\frac{\partial{\cal L}^{1\text{-loop}}_\text{HE}({\cal F})}{\partial{\cal F}}+2{\cal F}\xi\frac{\partial^2{\cal L}^{1\text{-loop}}_\text{HE}({\cal F})}{\partial{\cal F}^2}\frac{\chi+\ln(1-\chi)}{\chi^2}\,\kappa + {\cal O}(\kappa^2)
\end{equation}
and using this expression in \Eqref{eq:largeNcalLB}, up to linear order in $\kappa$ we obtain
\begin{align}
{\cal L}_{\rm HE}({\cal F})\big|_{N\to\infty}^{{\cal O}(\alpha^3)}&={\cal L}_{\rm HE}({\cal F})\big|_{N\to\infty}^{{\cal O}(\bar j_0^2)} +\frac{2}{3}{\cal F}\xi^3\biggl[{\cal F}\frac{\partial^2{\cal L}^{1\text{-loop}}_\text{HE}({\cal F})}{\partial{\cal F}^2}f_2(\chi) \nonumber\\
 &\hspace{1.5cm}+2\biggl({\cal F}\frac{\partial^2{\cal L}^{1\text{-loop}}_\text{HE}({\cal F}) }{\partial{\cal F}^2}+{\cal F}^2\frac{\partial^3{\cal L}^{1\text{-loop}}_\text{HE}({\cal F}) }{\partial{\cal F}^3}\biggr)f_3(\chi)\biggr]f_1(\chi)+{\cal O}(\alpha^4)\,.
 \label{eq:Lpert1}
\end{align}
Next, we divide by a factor of $\cal F$, define $\hat{\cal L}:={\cal L}_{\rm HE}^{1\text{-loop}}({\cal F})/\alpha$ and represent \Eqref{eq:Lpert1} as
\begin{align}
{\cal L}_{\rm HE}({\cal F})\big|_{N\to\infty}^{{\cal O}(\alpha^3)}/{\cal F}=
&-1+\alpha\bigl(\hat{\cal L}/{\cal F}\bigr)+2\alpha\xi\bigl(\partial_{\cal F}\hat{\cal L}\bigr)f_1(\chi) \nonumber\\
&-\xi^2\Bigl\{\Bigl[1-\alpha\bigl(\partial_{\cal F}\hat{\cal L}\bigr)\Bigr]f_2(\chi) -2\alpha\bigl({\cal F}\partial_{\cal F}^2\hat{\cal L}\bigr)f_3(\chi)\Bigr\} \nonumber\\
&+\frac{2}{3}\xi^3\Bigl\{\alpha\bigl({\cal F}\partial_{\cal F}^2\hat{\cal L}\bigr)f_2(\chi) +2\alpha\Bigl[\bigl({\cal F}\partial_{\cal F}^2\hat{\cal L}\bigr)+\bigl({\cal F}^2\partial_{\cal F}^3\hat{\cal L}\bigr)\Bigr]f_3(\chi)\Bigr\}f_1(\chi) \nonumber\\
&+{\cal O}(\alpha^5)\,,
 \label{eq:Lpert2}
\end{align}
where $\bigl(\hat{\cal L}/{\cal F}\bigr)\sim\bigl(\partial_{\cal F}\hat{\cal L}\bigr)\sim\bigl({\cal F}\partial_{\cal F}^2\hat{\cal L}\bigr)\sim\bigl({\cal F}^2\partial_{\cal F}^3\hat{\cal L}\bigr)\sim\alpha^0$.
Recalling the definition of the parameters  $\xi$ and $\chi$ in \Eqref{eq:xi+chi} and formally expanding these in powers of the fine-structure constant as 
\begin{equation}
\xi=\sum_{n=1}\xi_{(n)\,}\alpha^n\quad\text{and}\quad \chi=\sum_{n=1}\chi_{(n)\,}\alpha^n\,,
\end{equation}
we then  extremize \Eqref{eq:Lpert2} at each order in $\alpha$ for the expansion coefficients $\xi_{(n)}$ and $\chi_{(n)}$ this order depends on.
The corresponding conditions arising from the terms written explicitly in \Eqref{eq:Lpert2} fully determine the coefficients $\xi_{(1)}=-\bigl(\partial_{\cal F}\hat{\cal L}\bigr)$, $\chi_{(1)}=2\bigl({\cal F}\partial_{\cal F}^2\hat{\cal L}\bigr)$ and $\xi_{(2)}=-\bigl(\partial_{\cal F}\hat{\cal L}\bigr)^2$.
Upon plugging these values back into \Eqref{eq:Lpert2} we can extract the explicit expression for ${\cal L}_{\rm HE}({\cal F})\big|_{N\to\infty}$ up to cubic order in $\alpha$.
Apart from the Maxwell term and the one-loop Lagrangian, $-{\cal F}+{\cal L}_{\rm HE}^{1\text{-loop}}({\cal F})$, this yields
\begin{align}
{\cal L}_{\rm HE}({\cal F})\big|_{N\to\infty}^{{\cal O}(\alpha)}\,&=\frac{1}{2}{\cal F}\bigl[\partial_{\cal F}{\cal L}_{\rm HE}^{1\text{-loop}}({\cal F})\bigr]^2\,, \nonumber\\
{\cal L}_{\rm HE}({\cal F})\big|_{N\to\infty}^{{\cal O}(\alpha^2)}\,&=\frac{1}{2}{\cal F}\bigl[\partial_{\cal F}{\cal L}_{\rm HE}^{1\text{-loop}}({\cal F})\bigr]^2\biggl(\partial_{\cal F}{\cal L}_{\rm HE}^{1\text{-loop}}({\cal F})+\frac{4}{3}{\cal F}\partial_{\cal F}^2{\cal L}_{\rm HE}^{1\text{-loop}}({\cal F})\biggr)\,,  \nonumber\\
{\cal L}_{\rm HE}({\cal F})\big|_{N\to\infty}^{{\cal O}(\alpha^3)}\,&=\frac{1}{2}{\cal F}\bigl[\partial_{\cal F}{\cal L}_{\rm HE}^{1\text{-loop}}({\cal F})\bigr]^2\biggl(\bigl[\partial_{\cal F}{\cal L}_{\rm HE}^{1\text{-loop}}({\cal F})\bigr]^2+2\bigl[{\cal F}\partial_{\cal F}^2{\cal L}_{\rm HE}^{1\text{-loop}}({\cal F})\bigr]^2 \nonumber\\
&\hspace{2cm}+\frac{1}{9}\partial_{\cal F}{\cal L}_{\rm HE}^{1\text{-loop}}({\cal F})\bigl[31{\cal F}\partial_{\cal F}^2{\cal L}_{\rm HE}^{1\text{-loop}}({\cal F})+4{\cal F}^2\partial_{\cal F}^3{\cal L}_{\rm HE}^{1\text{-loop}}({\cal F})\bigr]\biggr)\,.
\label{eq:Lpert3}
\end{align}
As a consistency check we have verified that up to ${\cal O}(\alpha^2)$ exactly the same expressions are obtained when expanding \Eqref{eq:largeNcalLBquad2} in powers of the fine-structure constant making use of the scalings inferred in between Eqs.~\eqref{eq:chi0} and \eqref{eq:largeNcalLBquad2}.
On the other hand, upon subtracting the ${\cal O}(\alpha^3)$ contribution of \Eqref{eq:largeNcalLBquad2} from that in \Eqref{eq:Lpert3}, we obtain
\begin{equation}
 \Delta{\cal L}_{\rm HE}({\cal F})\big|_{N\to\infty}^{{\cal O}(\alpha^3)}=\frac{1}{18}{\cal F}^2\bigl[\partial_{\cal F}{\cal L}_{\rm HE}^{1\text{-loop}}({\cal F})\bigr]^3\biggl(7\partial_{\cal F}^2{\cal L}_{\rm HE}^{1\text{-loop}}({\cal F})+4{\cal F}\partial_{\cal F}^3{\cal L}_{\rm HE}^{1\text{-loop}}({\cal F})\biggr)\,,
 \label{eq:DeltaL}
\end{equation}
which should correspond to the expression for the second diagram in Fig.~\ref{fig:diags} (B).

From Eqs.~(3.14) and (3.15) of Ref.~\cite{Karbstein:2021gdi} it is clear that the contribution of this diagram to ${\cal L}_{\rm HE}({\cal F})\big|_{N\to\infty}$ can be represented as
\begin{align}
&\frac{1}{V^{(4)}}\frac{1}{3!}\int\ldots\int  \big(S_\psi^{(3)}[\bar A]\big)^{\sigma_1\sigma_2\sigma_3} D_{\sigma_1\mu}  \big(S_\psi^{(1)}[\bar A]\big)^{\mu} D_{\sigma_2\nu}  \big(S_\psi^{(1)}[\bar A]\big)^{\nu} D_{\sigma_3\rho}  \big(S_\psi^{(1)}[\bar A]\big)^{\rho} \nonumber\\
&=\frac{8}{3}\int_p\,(2\pi)^4\delta(p)\,\frac{p^\rho p^\gamma p^\alpha p^\tau}{(p^2)^2}\frac{\partial^3{\cal L}_{\rm HE}^{1\text{-loop}}(\bar F)}{\partial\bar F^{\rho\sigma}\partial\bar F^{\alpha\beta}\partial\bar F^{\mu\nu}}\frac{\partial{\cal L}_{\rm HE}^{1\text{-loop}}(\bar F)}{\partial\bar F^\gamma_{\ \ \sigma}}\frac{\partial{\cal L}_{\rm HE}^{1\text{-loop}}(\bar F)}{\partial\bar F^\tau_{\ \,\beta}}\frac{\partial{\cal L}_{\rm HE}^{1\text{-loop}}(\bar F)}{\partial\bar F_{\mu\nu}}\,, \nonumber\\
\intertext{which for ${\cal G}=0$ simplifies to}
&=\frac{1}{18}{\cal F}^2\bigl[\partial_{\cal F}{\cal L}_{\rm HE}^{1\text{-loop}}({\cal F})\bigr]^3\biggl(7\partial_{\cal F}^2{\cal L}_{\rm HE}^{1\text{-loop}}({\cal F})+4{\cal F}\partial_{\cal F}^3{\cal L}_{\rm HE}^{1\text{-loop}}({\cal F})\biggr)\,,
\end{align}
and thus indeed agrees with \Eqref{eq:DeltaL}.
This completes our consistency check in the perturbative domain.
It is interesting to see how in conjunction with a perturbative expansion the above extremization  allows to efficiently determine higher-order contributions to the Heisenberg-Euler effective Lagrangian at large $N$ without the need to work out explicitly any higher-order Feynman diagrams.

\section{Conclusions and Outlook}\label{sec:concls}

In the present work, we provided an explicit expression for the full constant-field Heisenberg-Euler effective Lagrangian ${\cal L}_{\rm HE}({\bar F})\big|_{N\to\infty}$ in the 't Hooft limit receiving contributions from arbitrary loop orders.
Our starting point was the formal expression for the effective action~\eqref{eq:largeNGamma} originally derived in Ref.~\cite{Karbstein:2021gdi}, which is a functional of the external field $\bar A^\mu$ and an auxiliary, IR divergent vector field $\bar j^\mu$. The latter is supposed to extremize this effective action and thus is implicitly also defined in terms of  $\bar A^\mu$.
In a first step, we analyzed the momentum structure of the vector field $\bar j^\mu$ and showed that in generic constant fields it can be parameterized by four constant scalar coefficients $\pi_p$ with $p\in\{T,\bar F\bar F,{}^\star\!\bar F{}^\star\!\bar F,{}^\star\!\bar F\bar F\}$. Making use of this fact, the extremization of the effective action for $\bar j^\mu$ can be traded for the much simpler extremization for these four scalars.
We then demonstrated that upon introducing the field strength tensor of the auxiliary field $\bar J^{\mu\nu}$, in constant fields the effective action~\eqref{eq:largeNGamma} can be cast into the form of \Eqref{eq:largeNGamma2}. This expression is fully determined by the renowned one-loop Lagrangian ${\cal L}_{\rm HE}^{1\text{-loop}}(\bar F)$. 
Using that in constant fields ${\cal L}_{\rm HE}^{1\text{-loop}}(\bar F)$ depends on $\bar F^{\mu\nu}$ only via the scalar invariants of the field $\cal F$ and $\cal G$, further simplifications are possible and the associated full effective Lagrangian ${\cal L}_{\rm HE}({\cal F},{\cal G})\big|_{N\to\infty}$ can be expressed in the form of \Eqref{eq:largeNcalL}. By extremizing this Lagrangian for the four scalar coefficients $\pi_p$, its superficial dependence on these unknowns can be completely eliminated. 
Subsequently, we focused on the special case where ${\cal G}=0$ and arrived at the rather compact expression~\eqref{eq:largeNcalLB} for ${\cal L}_{\rm HE}({\cal F})\big|_{N\to\infty}$ in this limit. This effective Lagrangian is characterized by just two scalar unknowns $\xi$ and $\chi$ to be eliminated by an extremization.
Finally, as consistency checks we explicitly demonstrated that our expression~\eqref{eq:largeNcalLB}  for ${\cal L}_{\rm HE}({\cal F})\big|_{N\to\infty}$ correctly reproduces (A) the all-order strong field limit studied in Ref.~\cite{Karbstein:2021gdi}, and (B) the perturbative result for ${\cal L}_{\rm HE}({\cal F})\big|_{N\to\infty}$ up to ${\cal O}(\alpha^3)$. 

As a natural continuation of the present work, in the future we in particular plan to numerically extremize the effective Lagrangian~\eqref{eq:largeNcalLB} for both scalars $\xi$ and $\chi$. This will allow us to study the fate of the full Heisenberg-Euler Lagrangian at large $N$ as a function of $\cal F$ up to arbitrarily large values of this parameter.
Because the Feynman diagrams that dominate the all-loop strong field limit~\cite{Karbstein:2019wmj} of standard $N=1$ flavor QED are precisely those surviving in the large $N$ limit studied in the present work, we expect this study to provide also important guidance for the strong-field behavior of standard external-field QED.

\acknowledgments

This work has been funded by the Deutsche Forschungsgemeinschaft (DFG) under Grant No. 416607684 within the Research Unit FOR2783/2.
I would like to thank Gerald V. Dunne for comments and inspiring discussions.

\end{document}